\documentclass[aps,pra,twocolumn,groupedaddress,showpacs]{revtex4-1}
\usepackage{graphicx}
\usepackage{amsmath}
\usepackage{textcomp}
\usepackage{mathrsfs}
\usepackage{amsfonts}
\usepackage{color}
\begin{document}

\newcommand{\vett}[1]{\mathbf{#1}}
\newcommand{\uvett}[1]{\hat{\vett{#1}}}
\newcommand{\beq}{\begin{equation}}
\newcommand{\eeq}{\end{equation}}
\newcommand{\bseq}{\begin{subequations}}
\newcommand{\eseq}{\end{subequations}}
\newcommand{\barr}{\begin{eqnarray}}
\newcommand{\earr}{\end{eqnarray}}
\newcommand{\bra}[1]{\langle #1|}
\newcommand{\ket}[1]{|#1\rangle}
\newcommand{\expectation}[3]{\langle #1|#2|#3\rangle}
\newcommand{\braket}[2]{\langle #1|#2\rangle}

\title{Harmonic generation in bent graphene with artificially-enhanced spin-orbit coupling}

\author{Arttu Nieminen$^1$}
\author{Marco Ornigotti$^1$}
\email{marco.ornigotti@tuni.fi}
\affiliation{$^1$Tampere University, Photonics Laboratory, Physics Unit, FI-33720 Tampere, Finland}

\begin{abstract}
We theoretically investigate the nonlinear response of bent graphene, in the presence of artificially-enhanced spin-orbit coupling, which can occur either via adatom deposition, or by placing the sheet of bent graphene in contact with a spin-orbit active substrate. We discuss the interplay between the spin-orbit coupling and the artificial magnetic field generated by the bending, for both the cases of Rashba and intrinsic spin-orbit coupling. For the latter, we introduce a spin-field interaction Hamiltonian addressing directly the electron spin as a degree of freedom. Our findings reveal that in this case, by controlling the amount of spin-orbit coupling, it is possible to significantly tune the spectrum of the nonlinear signal, achieving, in principle, efficient conversion of light from THz to UV region. 
\end{abstract}
\maketitle
 \section{Introduction}
Since its discovery in 2004 \cite{graphene1}, graphene has attracted a lot of interest in the scientific community, mainly due to its exquisite, and unexpected, electronic, mechanical, and thermal properties \cite{graphene2, graphene3}, but also fascinating optical properties, such as universal absorption \cite{absorbance}, ultrafast broadband response \cite{graphene4}, and large nonlinear optical responses \cite{graphene6,graphene7}, to name a few. Most of these properties derive directly from the presence, in the band structure of graphene, of Dirac cones, i.e., points in $k$-space, where valence and conduction band touch, thus giving rise to a gapless linear dispersion \cite{katsnelson_2012}. Monolayers of graphene also admit spin-orbit coupling (SOC), in the form of both intrinsic ($\Delta_I$) and Rashba ($\Delta_R$) coupling, the former originating as a true SOC due to the relativistic nature of electrons in graphene, while latter occurring only in the presence of an external electric field \cite{katsnelson_2012}. For the particular case of $\Delta_I>\Delta_R/2$, Kane and Mele discovered, for a finite monolayer of graphene, that the gap opened by SOC at the Dirac points sustains topologically protected edge states near its boundary \cite{SOC_ref}, where spin-dependent impurity backscattering is strongly suppressed, resulting in the so-called quantum spin Hall (QSH) edge states. The discovery of Kane and Mele, moreover, was so influential, that it gave birth to the exciting field of topological insulators \cite{topoInsulator}, which, soon after its discovery in the context of condensed matter physics, started contaminating other fields of physics, giving rise to new ideas, such as topological photonics \cite{topoPhotonics,topoPhotonics2}, topological mechanics \cite{topoOpto}, and topological atomic physics \cite{topoAtom}. 

The QSH effect, however, is quite hard to be observed experimentally in pristine graphene, since the value of the intrinsic SOC is too small, to allow its experimental verification \cite{soc1,soc2,soc3}. The QSH state, however, has been experimentally observed in other systems, such as HgTe quantum wells \cite{hgteWells}, InAs/GaSb quantum wells \cite{inasQW}, and WTe$_2$ \cite{wte2}, to name a few. To overcome the problem of small SOC in graphene, several different strategies have been proposed, ranging from increasing the Rashba coupling by depositing graphene on Ni surfaces \cite{rashba1,rashba2}, to a significant increase of the intrinsic SOC by adatom deposition of different compounds \cite{adatomSOC}, such as Indium, Thallium, or Bi$_2$Te$_3$ nanoparticles, the latter representing the first experimental evidence of the occurrence of QSH effect in graphene with artificially enhanced SOC \cite{adatomSOC2}. 

Contextually, several works investigated the effect of Rashba \cite{rashbaMag1,rashbaMag2,rashbaMag3} and intrinsic SOC on the electronic structure of graphene, in the presence of magnetic fields \cite{formalism}. The latter, in particular, have attracted considerable attention in the last years, because they can be realised by applying strain or bending to single and multilayered 2D materials \cite{extra1,extra2}, resulting in high artificial magnetic fields, which, contrary to the real ones, cannot break time-reversal symmetry. A comprehensive review on the topic can be found in Ref. \cite{reviewAGF}. 

Interestingly, though, only few works investigated the effects of (pseudo)magnetic fields in the nonlinear optical response of graphene, and they have been mainly focused on estimating how the third-order nonlinear susceptibility of graphene depends on the applied magnetic field, in the limit of strong magnetic field \cite{magNLO}. In a recent work, moreover, the role of a constant, out-of-plane magnetic field in the nonlinear response of graphene has been thoroughly investigated, revealing the possiblity to use the magnetic field strength to control the frequency conversion, up to the visible range \cite{ref1}. None of them, though, investigated the role SOC might have in shaping the nonlinear optical response of graphene, and, more generally, 2D materials. 

In this work, we investigate the effect of both intrinsic and Rashba SOC on the nonlinear signal generated by an ultrashort electromagnetic pulse impinging upon a flake of bent graphene. To do so, we extend the formalism recently developed by one of the authors in Ref. \citenum{ref1} to explicitly account for the presence of a nonzero SOC coupling, and accomodate explicitly spin dynamics in the model. For the case of intrinsic SOC, in particular, we exploit the fact that the spin degeneracy of the electronic bands is lifted and we introduce a spin-field interaction Hamiltonian and compare its action with the standard sub-lattice (i.e., minimal coupling) interaction Hamiltonian. Our results show, that by addressing directly the spin degree of freedom of Dirac electrons in graphene it is possible, by controlling the level of intrinsic SOC, to significantly broaden the spectrum of the nonlinear signal, allowing efficient conversion of light from THz to the UV region.

This work is organised as follows: in Sect. II we present the basic model used in this work, namely the Hamiltonian for a 2D material (graphene, in this specific case) in the presence of SOC. Section III is then dedicated to the case of bent graphene, and to deriving its eigenstates and eigenvalues, for both the cases of intrinsic and Rashba SOC. In Section IV, then, we briefly discuss how to introduce the spin-field interaction in the graphene Hamiltonian, and discuss explicitly the cases of linear and circular polarisation. Section V is then dedicated to the discussion of the nonlinear signal in the presence of SOC. Finally, conclusions are drawn in Sec. VI.
\section{Graphene Hamiltonian with SOC}
\begin{figure}[!t]
\centering
\includegraphics[width=0.5\textwidth]{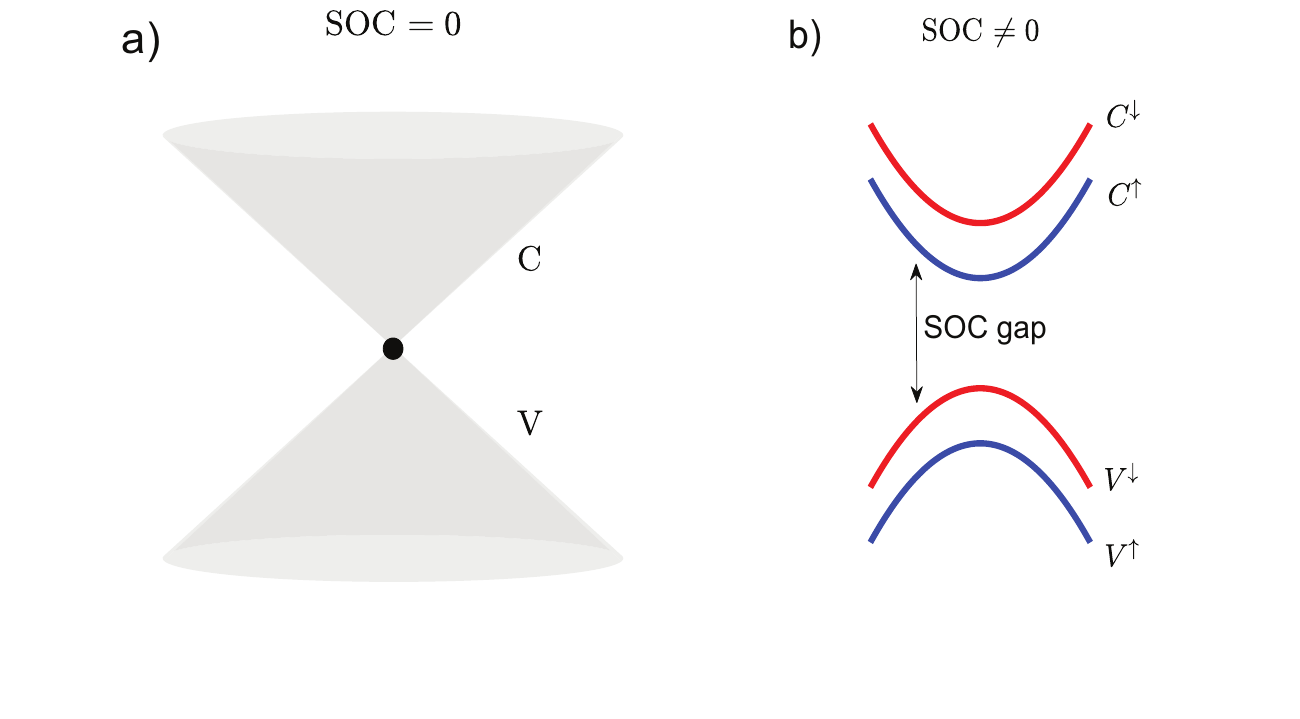}
\caption{\footnotesize \label{figure1} Band structure of graphene in the vicinity of one Dirac valley (for example, $K$), for both the case of no SOC (a), where the usual gapless, linear dispersion relation appears, and in the presence of SOC (b). Depending on the kind of SOC considered, a gap can be opened at the Dirac point (intrinsic SOC), and the spin-degeneracy of both valence and conduction bands can be lifted (Rashba SOC).}
\end{figure}

Electron dynamics in graphene are typically described using a 4-component spinor, $\Phi(\vett{r},t)=(\phi_K^A,\phi_K^B,\phi_{K'}^A,\phi_{K'}^B)^T$, where $\{A,B\}$ refer to the sub-lattice site (associated to the so-called pseudospin degree of freedom), referring to the two carbon atoms per unit cell, and the indices $\{K,K'\}$  indicate the two nonequivalent valleys in k-space \cite{katsnelson_2012}. If SOC is present, the spin degeneracy of the two Dirac bands in each valley is lifted, leading to a spin-resolved 4 band system in each valley (see Fig. \ref{figure1}). In this case, then, electron dynamics are completely described by means of a 8-component spinor $\Psi(\vett{r},t)\equiv (\Phi_{\uparrow}(\vett{r},t), \Phi_{\downarrow}(\vett{r},t))^T$, where $\{\uparrow,\downarrow\}$ is the spin index. To write the Hamiltonian for graphene in the presence of SOC, notice that $\Psi(\vett{r},t)$ depends on three independent degrees of freedom, namely pseudospin (sub-lattice), spin, and valley, each spanning one of three different two-dimensional subspaces, i.e.,  $\Psi(\vett{r},t)\in\mathcal{H}\equiv\mathcal{H}_{valley}\otimes\mathcal{H}_{AB}\otimes\mathcal{H}_{spin}$, where each individual subspace $\mathcal{H}_{\mu}$ is spanned by its own set of Pauli matrices $\sigma^{\mu}_{x,y,z}$, and $\text{dim}\{\mathcal{H}\}=8$. Since SOC does not mix the valley degree of freedom, we can factor out the valley degree of freedom and reorganise the elements of $\Psi(\vett{r},t)$ by introducing the spin-resolved valley spinor $\phi^{\xi}=(\phi_{\uparrow,A}^{\xi},\phi_{\uparrow,B}^{\xi},\phi_{\downarrow,A}^{\xi},\phi_{\downarrow,B}^{\xi})^T$, with $\xi=\{1,-1\}\equiv\{K,K'\}$ being the valley index, so that $\Psi(\vett{r},t)=(\phi^K,\phi^{K'})^T$. By doing so, we can then write the single-valley Hamiltonian in the presence of SOC as
\beq
\hat{H}^{\xi}=\hat{H}_0^{\xi}+\hat{H}_I^{\xi}+\hat{H}_R^{\xi},
\eeq
where
\beq\label{freeHam0}
\hat{H}_0^{\xi}=v_f\left(\xi\,p_x\,\mathbb{I}_s\otimes\sigma_x+p_y\,\mathbb{I}_s\otimes\sigma_y\right),
\eeq
is the free Hamiltonian, $\mathbb{I}_s$ is the identity matrix in spin subspace, $\sigma_{x,y}$ are the Pauli matrices spanning the two-dimensional sub-lattice space $\mathcal{H}_{AB}$, 
\beq\label{eq4}
\hat{H}_I^{\xi}=\xi\,\Delta_I\,s_z\otimes\sigma_z,
\eeq
(with $s_{x,y}$ being the Pauli matrices spanning the two-dimensional spin space $\mathcal{H}_{spin}$) describes intrinsic SOC, which accounts for the case where the electron spin is oriented perpendicular to
the graphene plane \cite{katsnelson_2012}, and
\beq\label{eq5}
\hat{H}_R^{\xi}=\Delta_R (\xi\,s_y \otimes \sigma_x  - s_x\otimes\sigma_y),
\eeq
is the Rashba Hamiltonian, describing the case in which the electron spin is oriented in the plane of graphene, and it is responsible for the spin-momentum locking \cite{katsnelson_2012}. 

In graphene, the magnitude of both SOC terms is generally small, with the Rashba term being $\Delta_R \simeq 10 \,\mu eV$  per V/nm \cite{Rashba}, when an electric field is applied, and the intrinsic SOC term $\Delta_I \simeq 24 \,\mu eV$ \cite{inSOC}. These effects, however, can be artificially enhanced by suitable adatom deposition, such as Indium and Thallium \cite{adatomSOC}, or by putting the graphene sheet in contact with tellurites-based nanoparticles \cite{adatomSOC2}. This results in an increase of the SOC of graphene of many orders of magnitude. 

It is worth noticing, that although in the remaining of this manuscript we will refer only to artificially enriched graphene, the model presented in this work can be easily adapted to any 2D material in the presence of SOC, such as, for example, transition metal dichalcogenides (TMDs).

\section{Bent graphene in the presence of SOC}
\begin{figure}[!t]
\centering
\includegraphics[width=0.5\textwidth]{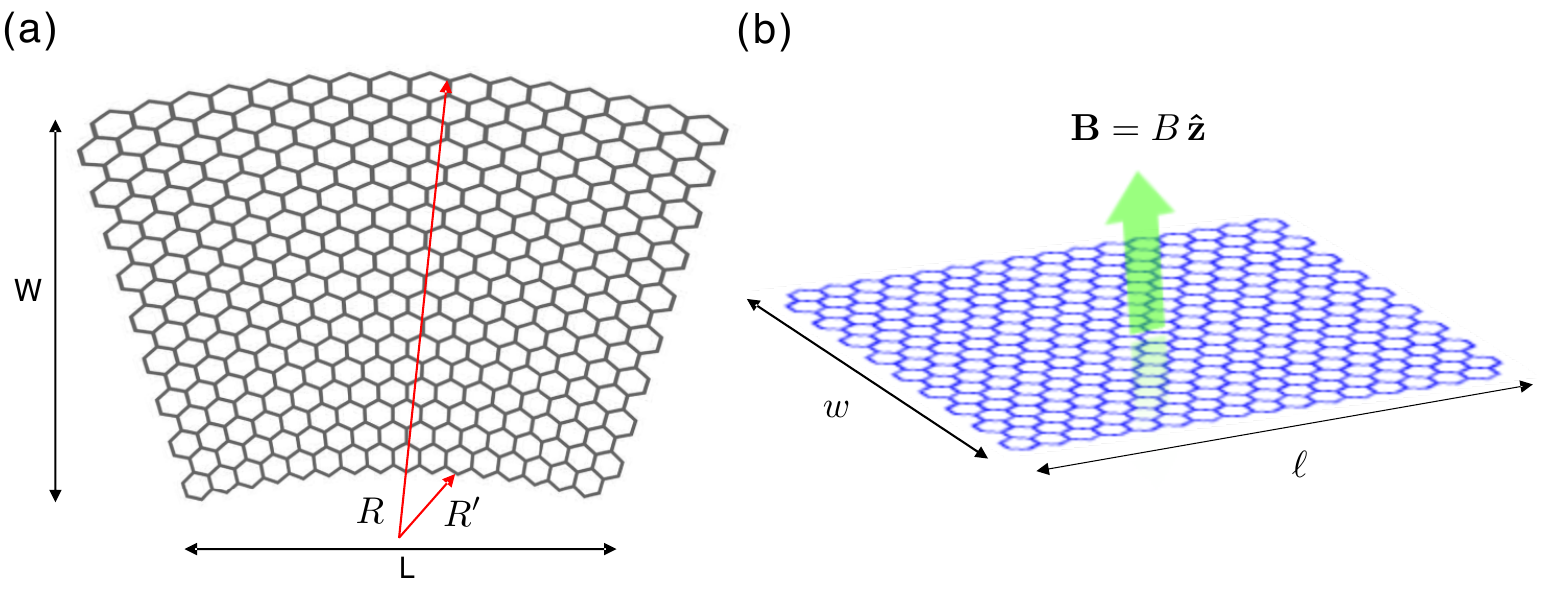}
\caption{\footnotesize \label{figure2} (a) Pictorial representation of a rectangular flake of grpahene, deformed into an arc. The radii of the upper, and lower edges are, respectively, $R$ and $R'$. (b) Flattened equivalent geometry of the bent graphene flake in panel (a). The curvature induced by the bending is replaced with an artificial gauge field $\vett{A}_s$, which gives rise to a uniform pseudomagnetic field $\vett{B}=B\,\uvett{z}$ parallel to the $z$-direction. The width $w$ and length $\ell$ of the flattened flake can be calculated from the bent structure, and might be extended to infinity for making calculations easier, as this operation does not change the essential role the pseudomagnetic field has in light-matter interaction.}
\end{figure}
The discussion above is valid for an unstrained single layer of graphene. When strain, or bending, is taken into account, an artificial gauge field (AGF) emerges, whose explicit expression depends on the nature of the strain or bending applied to the monolayer \cite{agf1}. In the absence of out-of-plane modulations, such induced AGF can be generally written as
\beq
\vett{A}^{(s)}=\pm C\left(u_{xx}-u_{yy}\right)\uvett{x}\mp 2Cu_{xy}\uvett{y},
\eeq
where $C$ is a suitable constant, and $u_{\mu\nu}=\partial_{\mu}u_{\nu}-\partial_{\nu}u_{\mu}$ is the strain tensor \cite{agf2}. Assuming a strain profile as the one described in Ref. \citenum{ref1}, and pictorially represented in Fig. \ref{figure2}, we obtain an AGF $\vett{A}^{(s)}=-B\,y\,\vett{x}$, corresponding to a uniform magnetic field oriented perpendicular to the graphene plane, i.e., $\vett{B}=B\,\uvett{z}$. The interaction of electrons in graphene with the AGF defined above can be introduced through minimal coupling, namely by replacing the kinetic momentum $\vett{p}$ of the electron, with the canonical momentum, i.e., $\vett{p}\mapsto \vett{p} + e \vett{A}^{(s)}$, in Eq. \eqref{freeHam0}. This is equivalent to introducing a magnetic interaction Hamiltonian 
\beq
\hat{H}_B^{\xi} =e\,v_F\,\xi\,A^{(s)}_x\,\mathbb{I}_s\otimes\sigma_x = -v_F\,e\,B\,y\,\xi\,\mathbb{I}_s\otimes\sigma_x,
\eeq
into Eq. \eqref{freeHam0}.
%
%
Since we are also considering the electron spin as a degree of freedom, simply adding the interaction term above to Eq. \eqref{freeHam0} is not enough, as we have to add an extra term to account for interaction of the electron spin with the synthetic magnetic field, i.e., a Zeeman term of the following form \cite{zeeman}:
\beq
\hat{H}_Z = \Delta_Z\,s_z\otimes\mathbb{I}_{AB},
\eeq
where $\mathbb{I}_{AB}$ is the identity matrix in sub-lattice space, and  $\Delta_Z = g_s\mu_B B/2$ (with $g_s$ being the gyromagnetic factor, and $\mu_B$ the Bohr magneton).

The total, single-valley Hamiltonian for SOC in the presence of an AGF is then given by $\hat{H}_{tot}^{\xi} = \hat{H}^{\xi} + \hat{H}_B^{\xi} + \hat{H}_Z^{\xi}$, and we can use it to the Dirac equation for electrons in the presence of SOC and under the action of an AGF as follows
\beq
i\hbar\frac{\partial}{\partial t} |\psi(\vett{r},t)\rangle = \sum_{\xi=\pm 1}\hat{H}_{tot}^{\xi}\, |\psi^{\xi}(\vett{r},t) \rangle.
\eeq
To solve the above equation, we write $\ket{\psi(\vett{r},t)}$ as as a linear combination of the instantaneous eigenstates of $\hat{H}_{tot}^{\xi}$, with time dependent expansion coefficients \cite{ishikawa}. To do so, despite $\hat{H}_{tot}^{\xi}$ admits closed-form eigenstates, in the form of parabolic cylinder functions for the general case of both instrinsic and Rashba SOC present in the system \cite{formalism}, in this work we discuss the two cases separately. This will provide a much easier framework, and will allow us to gain better insight on the role each of these two mechanisms has in the nonlinear optical response of graphene.
%
%
\subsection{Instantaneous eigenstates of $\hat{H}_{tot}^{\xi}$ for intrinsic SOC only}
To start with, we neglect Rasbha coupling, and set $\Delta_R=0$. As intrinsic SOC cannot lift the spin degeneracy of the bands near the Dirac point, but only introduces a nonzero gap proportional to $\Delta_I$, the two spin states\\ $\{\ket{\uparrow},\ket{\downarrow}\}$ can be treated independently, and the spin index $\beta=\pm 1\equiv\{\uparrow,\downarrow\}$ only enters as a parameter (and not as a quantum number) in the expression of the eigenstates and eigenvalues of $\hat{H}_{tot}^{\xi}$. Thanks to this, we can then write the single valley-single spin Hamiltonian as
\barr\label{Hint}
\hat{H}^{\xi}_{\beta}&=&v_f\left(\xi\,p_x\sigma_x+p_y\sigma_y-e\,B\,y\,\xi\,\sigma_x\right)\nonumber\\
&+&\beta\left(\xi\,\Delta_I\,\sigma_z+\Delta_Z\right),
\earr
and then $\hat{H}^{\xi}_{tot}=\hat{H}^{\xi}_{\uparrow}\otimes\hat{H}^{\xi}_{\downarrow}$, and therefore $\ket{\psi^{\xi}(\vett{r},t)}=\ket{\psi^{\xi}_{\uparrow}(\vett{r},t)}\otimes\ket{\psi^{\xi}_{\downarrow}(\vett{r},t)}$.

\begin{figure}[!t]
\centering
\includegraphics[width=0.45\textwidth]{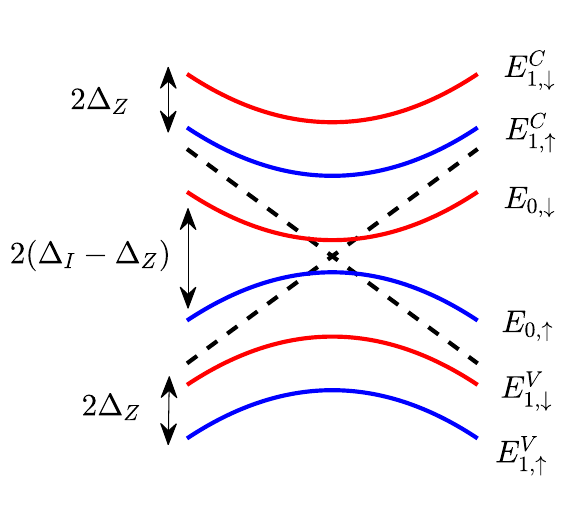}
\caption{\footnotesize \label{isoi_x} First few Landau energy levels in the vicinity of the Dirac point, in the presence of intrinsic SOC. As it can be seen, the main effect of the intrinsic SOC is to open a gap between valence and conduction bands (lifting the band degeneracy at the Dirac point). The lifting of the spin degeneracy of each level is due to the Zeeman coupling. The dashed lines, depicting the band structure of unperturbed, pristine graphene, have been added to help the visualisation of the band structure.}
\end{figure}

Written in the form above, it is easy to recognise Eq. \eqref{Hint} as a gapped Landau Hamiltonian, whose eigenvalues and eigenstates can be written in terms of harmonic oscillator eigenstates in the $y$ direction as \cite{landau}

\beq
\ket{\psi_{n,\beta}^{\xi}(y,t;p_x)}=\text{N}_{n,\beta}^{\xi}e^{i \left(\xi \frac{p_x}{\hbar} x-\frac{E_{n,\beta}^{\xi}}{\hbar} t\right)}\ket{\Phi_{n,\beta}^{\xi}(\eta)},
\eeq
with 
\beq
\ket{\Phi_{n,\beta}^{\xi,\alpha}(\eta)}=\hat{O}(\xi)\left(\begin{array}{c}
\nu_{n,\beta}^{\alpha}\,\phi_{n-1}(\eta)\\
\phi_n(\eta)
\end{array}\right),
\eeq
where $\hat{O}(\xi)=\mathbb{I}$ for $\xi=1$ (K valley), and $\hat{O}(\xi)=-\sigma_z\sigma_x$ for $\xi=-1$ (K' valley). Here, $\phi_n(\eta)$ are normalised one-dimensional harmonic oscillator eigenstates \cite{byron} (with $\phi_{-1}(\eta)=0$), $\eta = (-y + 2\ell_B^2 \xi p_x)/\ell_B$, (with $\ell_B=\sqrt{\hbar/eB}$ being the magnetic length),  $\text{N}_{n,\beta}^{\xi}$ is a normalisation constant,
\beq
\nu_{n,\beta}^{\alpha} = \frac{E_{n,\beta}^{\alpha} - E_0^{\alpha}}{\hbar \omega_c\sqrt{n}},
\eeq
and $E_{n,\beta}^{\alpha}=\alpha\hbar\omega_c\sqrt{n+(\Delta_I/\hbar\omega_c)^2}+\beta\Delta_z$ are the eigenvalues of Eq. \eqref{Hint} (with $\omega_c=v_F\sqrt{2}/\ell_B$ being the cyclotron frequency). $n\in\mathbb{N}_0$ indicates the Landau levels in the conduction ($\alpha=1$) and valence ($\alpha=-1$) band, respectively. Notice, moreover, that since the total Hamiltonian factors in spin space, there cannot be a common $n=0$ Landau level, as in the case of no intrinsic SOC \cite{ref1}, but rather two separate levels, with spin-dependent eigenvalues $E_{0,\beta}=-\beta(\Delta_I-\Delta_Z)$, and spin-polarized eigenstates

\beq
\ket{\Phi_{0,\beta}^{\xi}}=\left(\frac{1}{2\pi}\right)^{1/4}\left(\begin{array}{c}
0\\
\phi_0(\eta)
\end{array}\right),
\eeq
The corresponding band structure in the vicinity of the Dirac point for the first few Landau levels is shown schematically in Fig. \ref{isoi_x}.

\subsection{Instantaneous eigenstates of $\hat{H}_{tot}^{\xi}$ for Rashba SOC only}
We now turn our attention to the case, where only Rashba SOC is present, i.e., we set $\Delta_I=0=\Delta_Z$ in the total Hamiltonian. Since the Rashba coupling term $\Delta_R$ introduces a coupling between the spin states, we cannot factor $\hat{H}_{tot}^{\xi}$ in block diagonal form anymore. We then need to deal with the full four-dimensional Hamiltonian $\hat{H}_{tot}^{\xi}=\hat{H}^{\xi}+\hat{H}_R^{\xi}$. Its eigenstates and eigenvalues, however, can still be computed analytically and they can still be expressed in terms of harmonic oscillator states \cite{formalism} as follows
\beq
\ket{\psi_{n,\beta}^{\xi,\alpha}(y,t;p_x)}=\text{N}_{n,\beta}^{\alpha}e^{i\left(\xi \frac{p_x}{\hbar} x-\frac{E_{n,\beta}^{\alpha}}{\hbar}t\right)}\ket{\Sigma_{n,\beta}^{\alpha}(\eta)},
\eeq
where now $n\in\mathbb{Z}$, and
\beq
\ket{\Sigma_{n,\beta}^{\alpha}(\eta)}=\left(\begin{array}{c}
i\,n\,\Delta_R\, a_{n-1}\phi_{n-1}(\eta)\\
i\,\Delta_R\,E_{n,\beta}^{\alpha}\,a_{n}\phi_n(\eta)\\
\left[\left(E_{n,\beta}^{\alpha}\right)^2-n\right]\,a_{n}\phi_n(\eta)\\
\left[\frac{\left(E_{n,\beta}^{\alpha}\right)^2-n}{E_{n,\beta}^{\alpha}}\right]\,a_{n+1}\phi_{n+1}(\eta)
\end{array}\right),
\eeq
where $a_n = \sqrt{n!} \,$, and $E_{n,\beta}^{\alpha}$ are the eigenvalues of $\hat{H}^{\xi}_{tot}$, whose explicit expression is in general a complicated function of $\Delta_R$ and $\hbar\omega_c$ (see Ref. \citenum{formalism} for details). For the particular case $\varepsilon=\Delta_R/\hbar\omega_c\ll 1$, i.e., small Rashba coupling, however, we can find the following approximate analytical expression for the eigenvalues
%
%
\bseq \label{rashba_E}
\begin{align}
E_{n,+}^\alpha &= \alpha\hbar\omega_c\left(1+\frac{\varepsilon^2}{2}\right)\sqrt{n+1} + \mathcal{O} \left(\varepsilon^4 \right),\\
E_{n,-}^\alpha &= \alpha\hbar\omega_c\left(1-\frac{\varepsilon^2}{2}\right)\sqrt{n} + \mathcal{O} \left(\varepsilon^4 \right).
\end{align}
\eseq
Notice that in this limit $\hat{H}_{tot}^{\xi}$ admits two zero energy states, one corresponding to $n=0$ ($\ket{\Sigma_{0,-}^{\xi,\alpha}}$), and the other one to $n=-1$ ($\ket{\Sigma_{-1,+}^{\xi,\alpha}}$). 

Notice, moreover, that for $\Delta_R=0$, i.e., $\varepsilon=0$ in Eq. \eqref{rashba_E}, the condition $E_{m-1,+}^\alpha=E_{m,-}^\alpha$ holds, which implies a degeneracy of Landau levels.
This degeneracy is then lifted for $\Delta_R>0$, and in this case (and this case only) it makes sense to label the eigenstates with the index $\beta=\pm$, which, however, is not to be associated with a genuine spin index, since for the case of Rashba coupling, the bands are still degenerate in spin.

\begin{figure}[!t]
\centering
\includegraphics[width=0.45\textwidth]{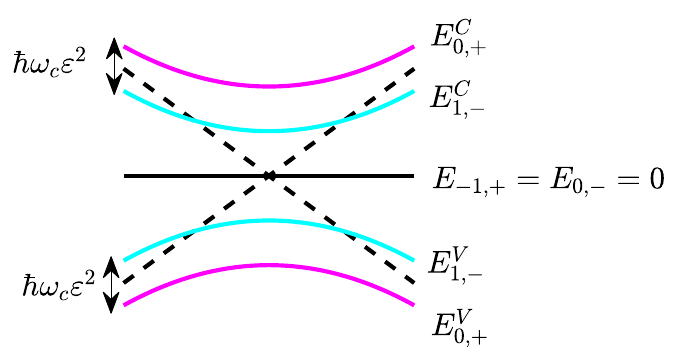}
\caption{\footnotesize \label{rashba_coup_x} Landau energy levels in the vicinity of the Dirac point, for the case of nonzero Rashba coupling, and for $n=\{0,1\}$.Contrary to the case of intrinsic SOC, in this case we observe the presence of a doubly degenerate zero state.The dashed lines, depicting the band structure of unperturbed, pristine graphene, have been added to help the visualisation of the band structure.}
\end{figure}

The band structure in the vicinity of the Dirac point for the case of nonzero (but small) Rashba coupling, including the two zero energy states, and the states corresponding to $n=\{0,1\}$ in both valence and conduction band, is reported in Fig. \ref{rashba_coup_x}.



\begin{figure*}[t!]
\centering
\includegraphics[width=0.9\textwidth]{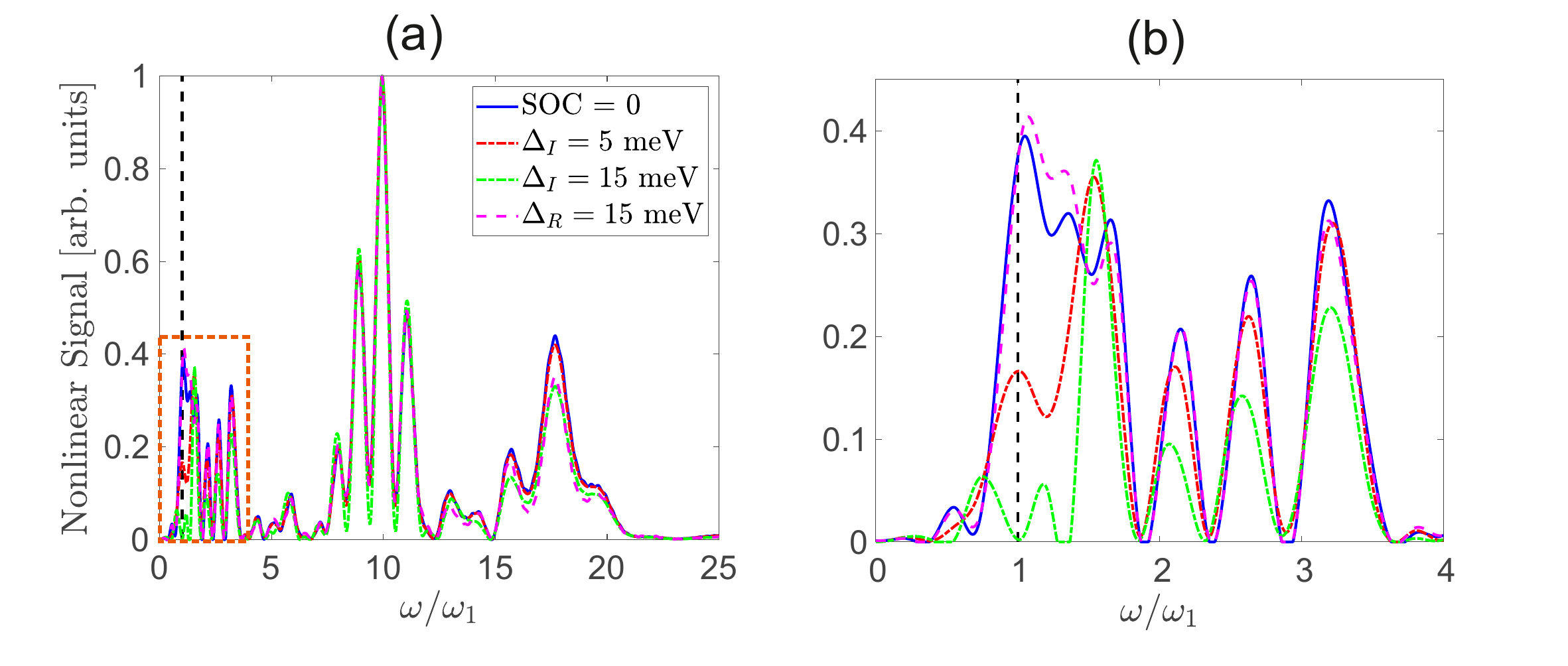}
\caption{\footnotesize \label{linear_plots_low} (a) Spectrum of emitted radiation by artificially enhanced bent graphene, for the case of impinging x-polarisation, and (b) zoom on the low-frequency part of the spectrum, i.e., $0\leq\omega\leq 4\omega_1$. The black, dashed line indicates the position of the fundamental frequency (i.e., $\omega=\omega_L=\omega_1$). The solid, blue line corresponds to the case of no SOC, and serves as a reference for the discussion. The red and green dashed lines represent, respectively, small, and high intrinsic SOC achievable with usual enhancement processes, while the pink, dashed line represents the same situation, but for the Rashba coupling. For these figures, the following parameters have been used: $\tau = 50$ fs, $\omega_L = 78 $ THz, $B = 2$ T, and $E_L=10^7$ V/m.}
\end{figure*}
\section{Interaction with the electromagnetic field}
We now consider the interaction of a flake of bent graphene in the presence of SOC, with an external electromagnetic field, and we then calculate its nonlinear response. To account for such an interaction, we employ minimal coupling, and simply make the electromagnetic vector potential $\vett{A}(t)$ appear via the minimal substitution $\vett{p}\mapsto \left(\vett{p}+e\vett{A}^{(s)}+e\vett{A}(t)\right)$, so that both the effects of an actual (electromagnetic pulse) and artificial (bending) gauge field are described within the same formalism. Throughout this whole section, we assume the electromagnetic vector potential to be written as $\vett{A}(t)=\mathcal{A}(t)e^{-i\omega_Lt}\uvett{f}+\text{c.c.}$, where $\uvett{f}$ is a suitable polarisation vector, $\omega_L$ is the pulse carrier frequency, and the pulse shape is assumed Gaussian, i.e., $\mathcal{A}(t)=E_L\tau\,\exp{\left(-(t-t_0)^2/\tau^2\right)}$, with $E_L$ being the pulse amplitude, and $\tau$ its duration. We moreover assume, for simplicity, that the electromagnetic field impinges normally on the bent graphene sheet. This assumption is justified by the fact, thatour model Hamiltonian for graphene automatically takes into account of a re-centering of $k$-space around the position of the Dirac points.

Accounting for all the aforementioned assumption, in this section we will therefore consider the following form of single valley Dirac equation
\beq \label{totHamLin}
i\hbar\frac{\partial}{\partial t} \ket{\Psi^{\xi}(x,y,t) } = \left[\hat{H}_{tot}^{\xi} +\hat{H}_{int}(t) \right] \ket{\Psi^{\xi}(x,y,t)},
\eeq
where the interaction term $\hat{H}_{int}(t)$ will assume different explicit forms, depending on the kind of interaction we are considering, as it is discussed below. 

In general, however, the interaction term $\hat{H}_{int}(t)$ can be gauged away by means of the phase transformation
\beq
\ket{\Psi^{\xi}(x,y,t)}=\int\,dp_x\,e^{i\frac{p_x}{\hbar} x}e^{-i\hat{G}(t)}\ket{\psi^{\xi}(y,t; p_x)},
\eeq
where $\hat{G}(t) = \hbar^{-1}\int_0^t \,ds\, \hat{H}_{int}(s)$. 
%
Following the reasoning from Ref. \cite{ref1}, we can calculate the instantaneous eigenstates of the equation above, which we define as $\ket{\Theta_{n,\beta}^{\xi,\alpha}(t)}$, for which we have
 $\ket{\Theta_{n,\beta}^{\xi,\alpha}(t)}=\ket{\Phi_{n,\beta}^{\xi,\alpha}(t)}$ for the case of intrinsic SOC, and $\ket{\Theta_{n,\beta}^{\xi,\alpha}(t)}=\ket{\Sigma_{n,\beta}^{\xi,\alpha}(t)}$ for the case of Rashba coupling.
 
To solve Eq. \eqref{totHamLin}, we then employ the Ansatz
\beq \label{sol_isoix}
\ket{\psi^{\xi}(y,t;p_x)} = e^{-i\hat{G}(t)} \sum_{n,\alpha,\beta} c_{n,\beta}^{\xi,\alpha}(t)\ket{\Theta_{n,\beta}^{\xi,\alpha}(y,t;p_x)},
\eeq
which amounts to considering the general electron dynamics for SOC bent graphene interacting with an external field as a weighted superposition of instantaneous Landau eigenstates. Substituting this Ansatz into Eq. \eqref{totHamLin} leads to solving a differential equation system for the expansion coefficients $c_{n,\beta}^\alpha(t)$:
\beq \label{c_eq}
\dot{c}_{m,\beta'}^{\xi,\alpha'}(t) =  i\sum_n\,\sum_{\alpha,\beta}\expectation{ \Theta_{m,\beta'}^{\xi,\alpha'}} {\hat{H}_{int}(t)}{ \Theta_{n,\beta}^{\xi,\alpha} } c_{n,\beta}^{\xi,\alpha}(t),
\eeq
where $\alpha,\beta=\{-1,1\}$. Notice, that the states $\ket{\Theta_{n,\beta}^{\xi,\alpha}}$ are orthogonal with respect to the index $n$, and the number of eigenstates involved in the sum above is regulated, essentially, by the initial conditions.
\begin{figure*}[t!]
\centering
\includegraphics[width=0.9\textwidth]{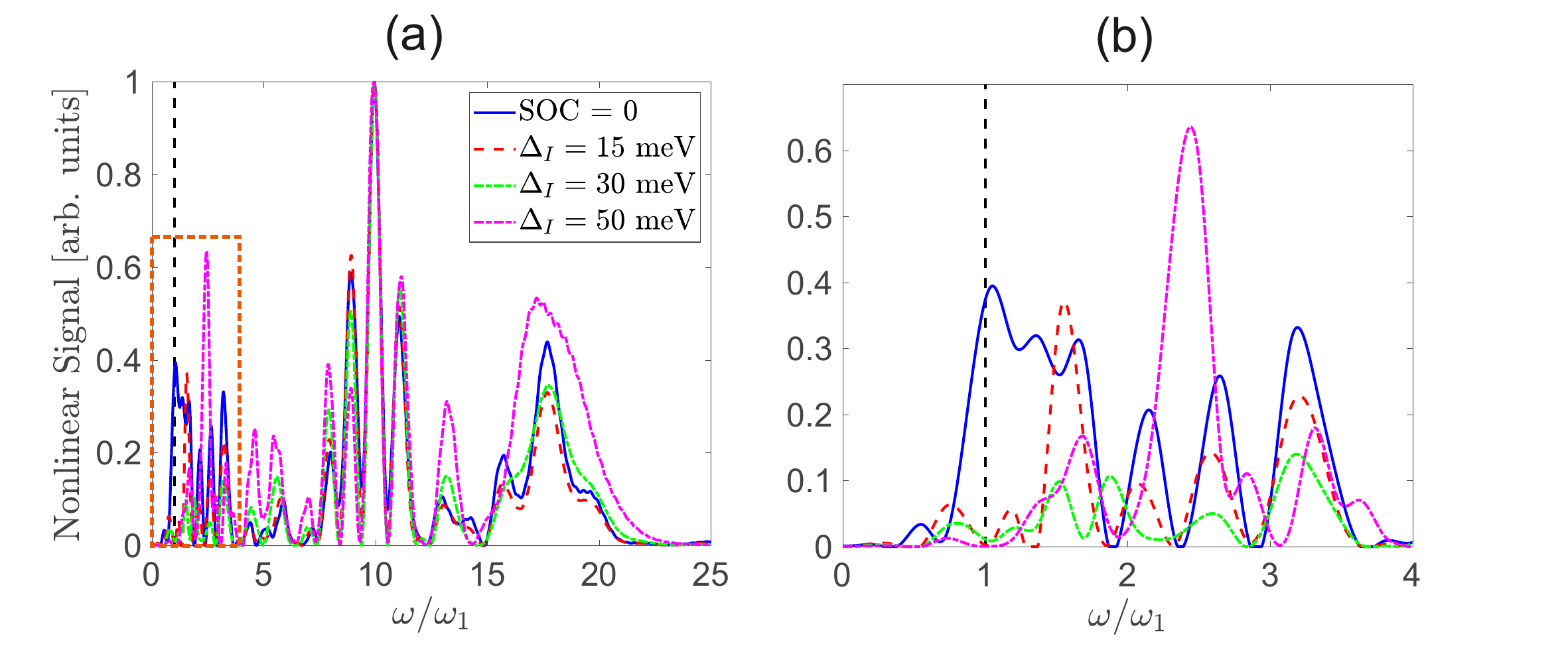}
\caption{\footnotesize \label{linear_plots_high} (a) Spectrum of emitted radiation by artificially enhanced bent graphene, for the case of impinging x-polarisation, and (b) zoom on the low-frequency part of the spectrum, i.e., $0\leq\omega\leq 4\omega_1$. The black, dashed line indicates the position of the fundamental frequency (i.e., $\omega=\omega_L=\omega_1$). The solid, blue line corresponds to the case of no SOC, and serves as a reference for the discussion. The dashed, red line corresponds to $\Delta_I=15$ meV, and depicts the same situation of the green, dashed line of Fig. \ref{linear_plots_low}. The dot-dashed green, and magenta lines correspond, respectively, to $\Delta_I=30$ meV and $\Delta_I=50$ meV. For these figures, the following parameters have been used: $\tau = 50$ fs, $\omega_L = 78 $ THz, $B = 2$ T, and $E_L=10^7$ V/m.}
\label{figure5}
\end{figure*}
We then consider two different regimes of light-matter interaction: first, we consider the usual sub-lattice coupling and assume that the impinging electromagnetic field is linearly polarised along the $x$-direction (namely, the one indicated by $\ell$ in Fig. \ref{figure2}), i.e., we choose $\uvett{f}=\uvett{x}$. The minimal coupling Hamiltonian in this case reads
\beq \label{intHamLin}
\hat{H}_{int}(t)=e\,v_F\,\mathbb{I}_s\,\otimes\,\xi\,A_x(t)\,\sigma_x.
\eeq

Secondly, we introduce a spin-field interaction term, which couples the photon and electron spin directly, thus granting us access to spin dynamics. To do so, we consider the case of a circularly polarised pulse, for which we choose $\uvett{f}=\uvett{h}_{\lambda}$, where $\uvett{h}_{\lambda}=\left(\uvett{x} + i \lambda \uvett{y} \right)/\sqrt{2}$ is the helicity basis \cite{mandelWolf}, and $\lambda=\pm 1$ is the photon spin angular momentum (SAM), corresponding to left-handed ($\lambda=+1$) or right-handed ($\lambda=-1$) circular polarisation. In this case, since the impinging pulse is carrying SAM, we can introduce an extra interaction Hamiltonian that describes the SAM-electron spin interaction, rather than the usual sub-lattice interaction described by Eq. \eqref{intHamLin} of the form
\beq \label{intHamCirc}
\hat{H}_{int}^{SF}(t) = e\,v_F\,A_\mu(t)\, s_\mu \otimes \mathbb{I}_{AB},
\eeq
where the superscript $SF$ stands for \emph{spin-field}, to emphasise the nature of the interaction, and distinguish the above interaction Hamiltonian from Eq. \eqref{intHamLin}.

Notice, that $\hat{H}^{SF}_{int}(t)$ makes only sense as interaction Hamiltonian when the electron spin can be addressed as an independent degree of freedom, namely only in the intrinsic SOC case, with nonzero Zeeman effect. In all other cases, where the spin degeneracy is not lifted, the interaction Hamiltonian cannot be written in this form, as one couldn't define Pauli matrices for the spin states, since spin is not an available degree of freedom.

\subsection{Dirac Current and Nonlinear Signal}
The nonlinear response of graphene can be estimated by first calculating the electric current generated by the interaction of the electromagnetic field with the graphene layer, namely
\beq
J_{\mu}^{\xi}(t)=\int\,dx\,dy\,\expectation{\Psi^{\xi}}{\hat{\mathcal{J}_{\mu}}}{\Psi^{\xi}},
\eeq
where $\hat{\mathcal{J}_{\mu}}$ is a suitable current operator, whose definition depends on the kind of interaction described above, namely $\hat{\mathcal{J}_{\mu}}=\mathbb{I}_s\otimes\sigma_{\mu}$ for the interaction described by Eq. \eqref{intHamLin}, and $\hat{\mathcal{J}_{\mu}}=s_{\mu}\otimes\mathbb{I}_{AB}$ for the spin-field interaction described by Eq. \eqref{intHamCirc}.
From the Dirac current, we can then calculate the spectrum of the emitted radiation, the so-called nonlinear signal, as a function of frequency, as
\beq \label{radiation}
I(\omega) \propto |\omega\,\Tilde{\vett{J}}(\omega)|^2,
\eeq
where $\Tilde{\vett{J}}(\omega)$ is the Fourier transform of the Dirac current $\vett{J}(t)$.

\section{Nonlinear signal}
We now have all the tools needed to investigate the nonlinear response of bent graphene, in the presence of artificially enhanced SOC, considering the effects of both the sub-lattice and spin-field interactions. We assume a pseudomagnetic field of magnitude $B=2$ T, and an impinging electromagnetic pulse with amplitude $E_L=10^7$ V/m, and duration of $\tau=50$ fs, with a carrier frequency of $\omega_L=78$ THz, fully resonant with the transition from between the lowest- and the zeroth state in the case of no SOC, and nearly resonant between the lowest and second-lowest states in the case when SOC is present. As initial conditions, we assume, for both the cases of intrinsic and Rashba SOC,  $\vett{c}^T(0)=(1\,0\,0\,1\,0\,0)^T/\sqrt{2}$. For the case of intrinsic SOC, this initial condition corresponds to assuming equal population of the lowest spin-up and spin-down states, i.e., $\vett{c}^T(0)=\left(c_{1,\uparrow}^V(0)\, c_{0,\uparrow}(0)\, c_{1,\uparrow}^C(0)\, c_{1,\downarrow}^V(0)\, c_{0,\downarrow}(0),c_{1,\downarrow}^C(0)\right)^T$. For the case of Rashba coupling on the other hand, since spin is not a viable quantum number anymore, the initial condition above just reduces to assuming equal population in the $+$ and $-$ states of valence band, since $\vett{c}^T(0)=\left(c_{1,-}^V(0)\, c_{0,-}(0)\, c_{1,-}^C(0)\, c_{0,+}^V(0)\, c_{-1,+}(0)\, c_{0,+}^C(0)\right)^T$.
\begin{figure}[t!]
\centering
\includegraphics[width=0.5\textwidth]{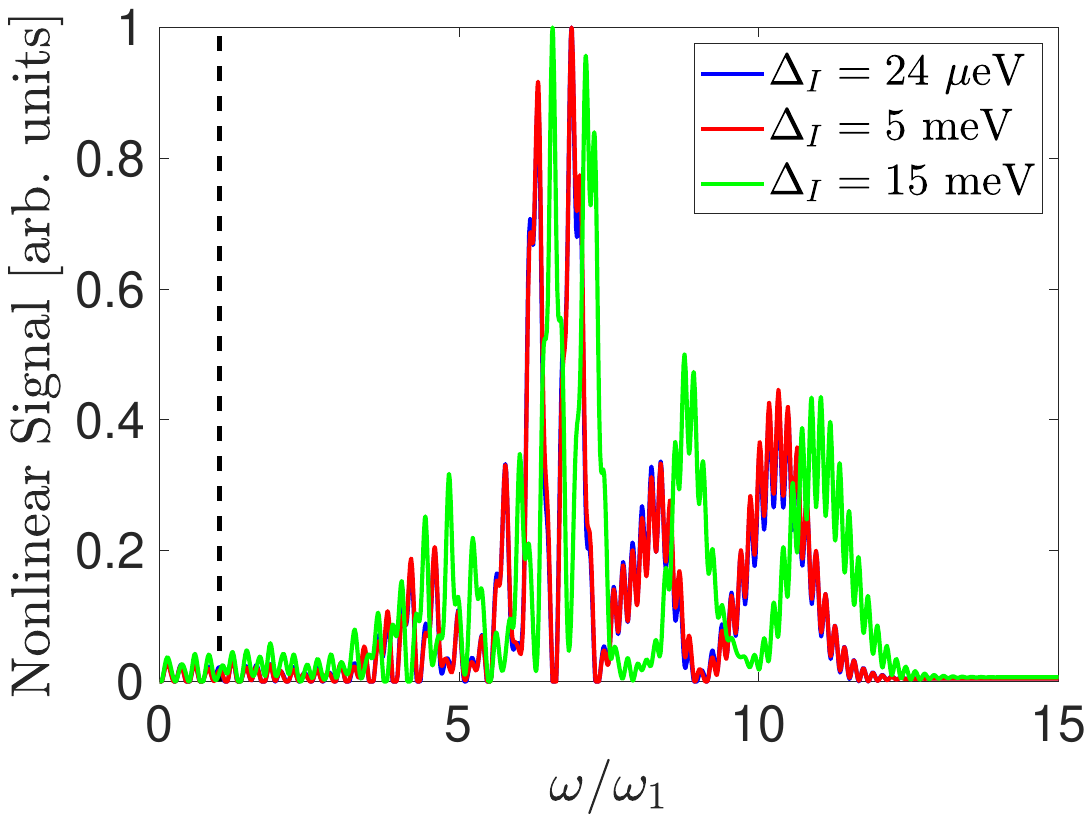}
\caption{\footnotesize \label{circular_spec_low} Spectrum of emitted radiation in the case of left-handed circularly polarised impinging pulse, for the case of low intrinsic SOC values. The black, dashed line indicates the position of the fundamental frequency (i.e., $\omega=\omega_L=\omega_1$). For these figures, the following parameters have been used: $\tau = 50$ fs, $\omega_L = 160$ THz, $B = 2$ T, and $E_L=10^7$ V/m.}
\end{figure}
\subsection{Linear Polarisation}
We start by first discussing the case of sub-lattice coupling with an $x$-polarised impinging field. To this aim, we mainly consider the effect of intrinsic ($\Delta_I$) SOC, as, as we will show, the influence of Rashba ($\Delta_R$) SOC is not so significant. To corroborate this statement, we have first performed simulations using typical values of both intrinsic and Rashba SOC, for the case of artificially-enhanced graphene, and chose $\Delta_I=5$ meV, and $\Delta_R=15$ meV. The result of these simulations is reported in Fig. \ref{linear_plots_low}. As it can be seen from panel (a), the presence of SOC of any kind does not drastically change the nonlinear response of bent graphene, but rather introduces small changes, mainly in the spectral region around the fundamental frequency, and in the high-harmonics region.

Let us first discuss the impact of the Rashba coupling (pink line in Fig. \ref{linear_plots_low}). We see from Fig. \ref{linear_plots_low}(b), that the main spectral region where Rashba SOC affects the nonlinear signal is the low-frequency region, around the fundamental frequency $\omega_L=\omega_1$). In this region, in fact, the nonlinear spectrum near the fundamental is slightly deformed. Overall, however, Fig. \ref{linear_plots_low} clearly shows how even a big chosen value of $\Delta_R=15$ does not introduce significant changes in the nonlinear spectrum. This observation then allows us to conclude, that Rashba SOC does not really contribute significantly to the nonlinear signal, and cannot therefore be used as an active control parameter to shape and engineer the nonlinear response of bent graphene.

On the other hand, we clearly see from Fig. \ref{linear_plots_low} (b), that the intrinsic SOC has a much higher impact on the nonlinear response of bent graphene. In fact, although from panel (a) we can see that a significant change in magnitude of the SOC from $\Delta_I=5$ meV (dashed, red line) to $\Delta_I=15$ meV (dashed, green line) does not have great impact on the high frequency side of the spectrum, where it only introduces a slight redistribution of energy [see, for example, the $\Delta_I$-dependent peak modulation around the 18th harmonic in panel (a)], it has quite a significant impact on the low frequency part of the nonlinear signal. A careful investigation of panel (b), in fact, reveals, that although the usual harmonic-oscillator-like structure of low harmonics \cite{ref1} is preserved also in the case of SOC, higher values of $\Delta_I$ tend to redistribute energy around the 2nd harmonic in a more efficient way than lower (or absent) values of SOC. For $\Delta_I=15$ meV, in particular, we see the emergence of a train of almost equally spaced non-integer harmonics, covering the range $[\omega_L,4\,\omega_L]$. 

This observation sparks the interesting question, whether this trend can be pushed forward, and intrinsic SOC can be used to effectively control the shape of the nonlinear response, at least in some frequency region. If this would be the case, artificially enhancing graphene could be a viable choice to tune its optical properties, even in ``real time". 

To this aim, in Fig. \ref{linear_plots_high} we compare the nonlinear signal produced by artificially enhanced graphene with progressively increasing values of intrinsic SOC, i.e., $\Delta_I=15$ meV (dashed, red line in Fig. \ref{linear_plots_high}), $\Delta_I=30$ meV (green, dot-dashed line in Fig. \ref{linear_plots_high}), and $\Delta_I=50$ meV (magenta, dot-dashed line in Fig. \ref{linear_plots_high}). As it can be seen, increasing the intrinsic SOC has different effects in different parts of the nonlinear spectrum. For high harmonics ($\omega\geq 15\omega_1$), increasing the amount of SOC in the system has the direct result of progressively broadening the harmonic peaks, until, for $\Delta_I=50$ meV, we obtain a single, broad peak, with higher intensity than the initial harmonics (blue line in Fig. \ref{linear_plots_high}). For the range $5\omega_1\leq 15\omega_1$, on the other hand, no significant changes seem to appear, as the basic structure of the nonlinear spectrum remains essentially the same, up to a small redistribution of energy between the harmonics. The situation for frequencies, i.e., $\omega\leq 5\omega_1$, however, is quite different, as can be seen from Fig. \ref{linear_plots_high} (b). While for small values of $\Delta_I$ the characteristic equally-spaced spectrum is still visible, although slightly red-shifted, for high values of $\Delta_I$ the situation changes drastically, as the harmonic-oscillator-like peaks, typical of the nonlinear response in this region \cite{ref1}, disappears, and a single, intense peak appears at approximately the non-integer frequency $\omega=5\omega/2$. Moreover, comparing the blue curve (no SOC) with all the others reveal how the presence of SOC makes the peak at the fundamental frequency $\omega=\omega_1$ disappear, meaning that SOC encourages a total redistribution of energy from the pump pulse to the different harmonics created in the material. This is an indication on how SOC can be used to tune the nonlinear response of artificially enhanced graphene, to generate devices for efficient conversion of light to specific frequencies, that do not need being integer multiples of some fundamental frequency.

To conclude this subsection, we would like to point out, that although this situation is highly unlikely to be observed in graphene, where even with artificial enrichment the SOC values achievable still remain in the meV regime, our results could be useful to describe the effect of SOC on the nonlinear signal of 2D materials in general, whose electronic properties can be described by a graphene-like Hamiltonian. TMDs, in particular, are a good example of that, since SOC is already quite big in such materials, and it could be enhanced even further with techniques similar to those utilised for graphene.
\subsection{Circular Polarisation}
\begin{figure*}[t!]
\centering
\includegraphics[width=1\textwidth]{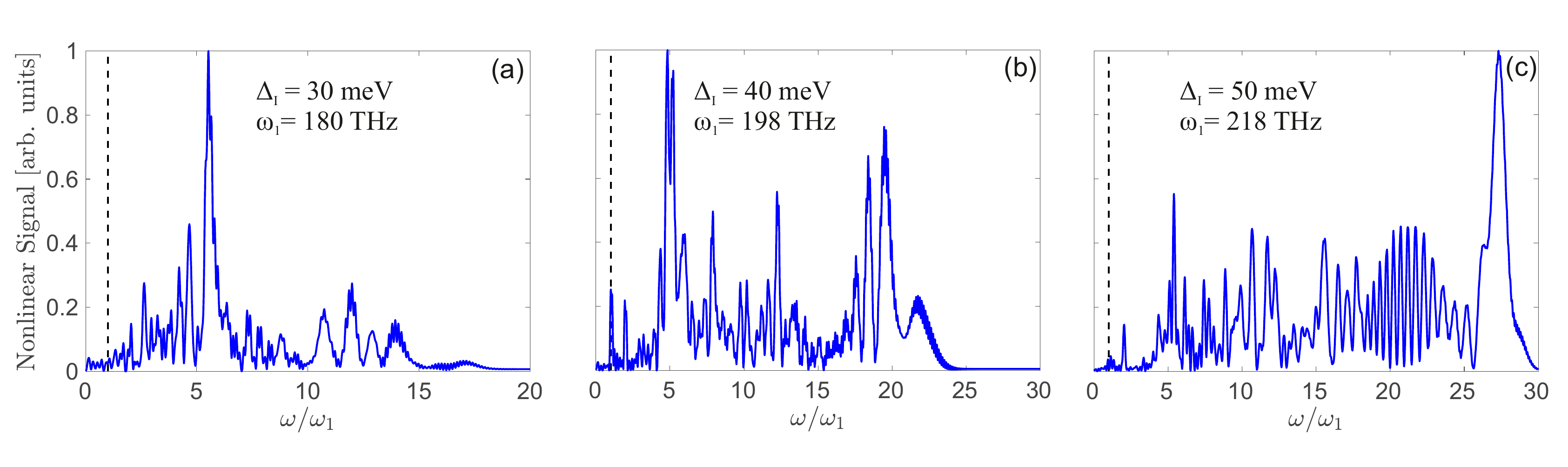}
\caption{\footnotesize \label{circular_spec_high} Spectrum of emitted radiation in the case of left-handed circularly polarised impinging pulse, for the case of high SOC values. The laser frequency $\omega_L$ has been adjusted to the various levels of intrinsic SOC, to ensure that the impinging pulse has no detuning. Notice, how the spectrum broadens with increasing values of $\Delta_I$, reaching, in panel (c), almost the 30th harmonic. The black, dashed line indicates the position of the fundamental frequency (i.e., $\omega=\omega_L=\omega_1$). For these figures, the following parameters have been used: $\tau = 50$ fs, $B = 2$ T, and $E_L=10^7$ V/m.}
\end{figure*}

We now turn our attention to the spin-field interaction Hamiltonian and investigate what kind of nonlinear signal such an interaction reveals. Here, as we did in the previous subsection, we only concentrate on intrinsic SOC, and discuss the cases for both small and large values of $\Delta_I$, whose results are depicted in Figs. \ref{circular_spec_low} and \ref{circular_spec_high}, respectively. In both figures, the laser parameters are the same that those used for the case of linear polarisation, except the carrier frequency, which in Fig. \ref{circular_spec_high} changes with increasing SOC,  to match the different values of SOC used for the simulations, so that the impinging electromagnetic pulse will always be resonant with an actual transition between a level in the valence band, and one in the conduction band, and no detuning will be present. 

For small values of $\Delta_I$, as it can be seen from Fig. \ref{circular_spec_low}, the situation is similar to the case of linear poarisation, and no appreciable changes can be observed. However, for values of $\Delta_I\geq 15$ meV (see the green line in Fig. \ref{circular_spec_low}), we start observing a significant blue shift of the nonlinear response. Notice, moreover, that the blue shift is not constant over the whole spectrum, but it is chirped in such a way, that lower frequencies experience a smaller shift, than the higher ones. This blue shift can be interpreted, with the help of Fig. \ref{isoi_x}, as the progressive increase of the gap between the (still equally spaced) Landau levels in valence and conduction bands, which become bigger as $\Delta_I$ increases.

If we increase $\Delta_I$ even further, as shown in Fig. \ref{circular_spec_high}, we witness a progressive broadening of the harmonic spectrum and the possibility, at very high SOC values [see Fig. \ref{circular_spec_high} (c)], of efficiently generating high harmonics. In particular, in the case shown in Fig. \ref{circular_spec_high} (c), we can excite the 28th harmonic of the fundamental frequency $\omega_L=217$ THz, corresponding to a wavelength of about $\lambda_{28}=310$ nm, well within the UV region of the electromagnetic spectrum. 
\section{Conclusions}
In this work, we theoretically investigated the effect of spin-orbit coupling (SOC) on the nonlinear response of a sheet of graphene under the action of an artificial magnetic field, for both the cases of Rashba and intrinsic SOC. The latter, in particular, allows for direct spin-field coupling, via the Hamiltonian in Eq. \eqref{intHamCirc}, which shows interesting features.

Our results show, that although for realistic values of both intrinsic and Rashba SOC in graphene no significant changes are induced in the nonlinear signal by SOC, controlling the amount of intrinsic SOC in 2D materials in the presence of an artificial magnetic field can result in a broadening of the spectrum of harmonics, and can even allow efficient conversion of light from the THz to the UV regions of the electromagnetic spectrum. This might lead to novel ways of generating spatially-varying frequency generation devices based on 2D materials, which could be achieved, for example, by inhomogeneously depositing SOC-active compounds over the surface of the 2D material, thus de-facto creating a gradient of intrinsic SOC through the material surface.
\section*{Acknowledgements}
The authors acknowledge the financial support of the Academy of Finland Flagship Programme (PREIN - decisions 320165). 
\bibliography{references}
\end{document}